\documentclass{elsart3}
\usepackage{graphicx,amssymb}
\journal{Canadian Journal of Physics}
\begin{document}
\begin{frontmatter}

\title{Laser spectroscopy of hyperfine structure in highly-charged ions: \\a test of QED at high fields}
\author{D.F.A. Winters\corauthref{cor}},
\corauth[cor]{Corresponding author.} \ead{d.winters@gsi.de}
\author{M. Vogel}
\address{GSI mbH, Planckstrasse 1, Darmstadt D-64291, Germany}
\author{D.M. Segal, R.C. Thompson}
\address{Blackett Laboratory, Imperial College, Prince Consort Road, London SW7 2BW, United Kingdom}
\author{W. N\"ortersh\"auser}
\address{Universit\"at Mainz, Fritz-Strassmann-Weg 2, Mainz D-55099, Germany}
\address{GSI mbH, Planckstrasse 1, Darmstadt D-64291, Germany}

\begin{abstract}
An overview is presented of laser spectroscopy experiments with
cold, trapped, highly-charged ions, which will be performed at the
HITRAP facility at GSI in Darmstadt (Germany). These
high-resolution measurements of ground state hyperfine splittings
will be three orders of magnitude more precise than previous
measurements. Moreover, from a comparison of measurements of the
hyperfine splittings in hydrogen- and lithium-like ions of the
same isotope, QED effects at high electromagnetic fields can be
determined within a few percent. Several candidate ions suited for
these laser spectroscopy studies are presented.
\end{abstract}

\begin{keyword}
QED, highly-charged ions, hyperfine structure, laser spectroscopy,
trapping and cooling \PACS 12.20.Fv, 21.10.Ky, 32.10.Fn, 32.30.-r
\end{keyword}
\end{frontmatter}

\section{Introduction}
Quantum electrodynamics (QED) was the first quantum field theory
to be formulated and has successfully passed every experimental
test at low and intermediate fields. A well-known example of QED
effects at low fields ($\sim 10^9$ V/cm) is the Lamb shift in
hydrogen \cite{LAM47}. At low fields, the QED effects (self-energy
and vacuum polarisation) can still be treated as a perturbation,
only taking into account lower order terms \cite{BEI00}. However,
up to now QED calculations have never been tested at high fields
($\sim 10^{15}$ V/cm) because such fields cannot be produced in a
laboratory, nor by the strongest lasers available. At high fields,
perturbative QED is no longer valid and higher order terms become
important as well \cite{BEI00}. Experiments carried out at high
fields therefore test different aspects of QED calculations and
are complementary to high-precision tests of the lower order
terms.

Heavy atoms that have been stripped of almost all their electrons,
the so-called highly-charged ions (HCI), are ideal `laboratories'
for tests of QED at high fields. These ions have, for example,
electric field strengths of the order of $10^{15}$ V/cm close to
the nucleus \cite{BEI00} and can be produced at high velocities at
the Gesellschaft f\"ur Schwerionenforschung (GSI) in Darmstadt,
Germany.

At the HITRAP facility, which is currently being built at GSI,
ions coming from the experimental storage ring (ESR) with MeV
energies will be slowed down by linear and radiofrequency stages
to keV kinetic energies, trapped and cooled down to sub-eV
energies, and finally made available for experiments. Within the
HITRAP project, instrumentation is being developed for
high-precision measurements of atomic and nuclear properties, mass
and g-factor measurements and ion-atom and ion-surface interaction
studies \cite{QUI01,BEI05,KLU05}.

\section{Hydrogen- and lithium-like ions}
Hydrogen- and lithium-like ions are the best candidates for our
studies, since they have $s$-electrons which are very close to the
nucleus. The (higher order) QED effects are most pronounced at the
high fields close to the nucleus, therefore the best measurable
quantity is the ground state hyperfine splitting (HFS). Due to the
simple electronic structure of H- and Li-like species, accurate
(higher order) calculations of ground state HFS can be done, which
will then be compared with accurate experimental results.

As a first approximation, good within about 4\%, the energy of the
$(1s)~^2$S$_{1/2}$ ground state HFS of hydrogen-like ions is given
by \cite{BEI00,SHA94}:
\begin{equation}
\label{eq1} E_{HFS}=\alpha (Z \alpha)^3 g_I \frac{m_e}{m_p}
\frac{2 (2I+1)}{3} m_e c^2 A_s (1-\delta)
\end{equation}
where $\alpha$ is the fine structure constant, $g_I=\mu /(\mu_N
I)$ is the nuclear $g$-factor (with $\mu$ the nuclear magnetic
moment and $\mu_N$ the nuclear magneton), $I$ the nuclear spin,
$m_e$ and $m_p$ are the electron and proton mass, respectively,
and $c$ is the speed of light. Equation (\ref{eq1}) represents the
normal ground state HFS multiplied by a correction $A_s$ for the
relativistic energy of the $s$-electron, and by a factor
$(1-\delta)$, which takes the `Breit-Schawlow' (BS) effect into
account. The BS effect is due to the spatial distribution of the
nuclear charge. It corrects for the fact that we cannot assume a
homogeneous charge distribution over a spherical nucleus. The
values for $\delta$ were taken from \cite{SHA94}, those for $g_I$
and $I$ from \cite{FIR98}. In principle eq.(\ref{eq1}) should also
contain a correction for the finite nuclear mass, but since this
correction is very small it can be neglected \cite{BEI00}. The
energy of the $(1s^2 2s)~^2$S$_{1/2}$ ground state HFS of
lithium-like ions only differs from eq.(\ref{eq1}) by a factor
$1/n^3=1/8$ and by the $A_s$-value \cite{BEI00}.

However, eq.(\ref{eq1}) requires two further important
corrections, the one of most interest to us being that which
corrects for the QED effects. The other correction takes the
`Bohr-Weisskopf' (BW) effect into account \cite{BOH50}. The BW
effect is due to the spatial distribution of the nuclear
magnetisation and is only known with an accuracy of 20-30 \%,
which is mainly due to the single-particle model used for its
calculation \cite{SHA97}. Unfortunately, the QED effects are of
the same order of magnitude as the uncertainty in the BW effect
\cite{SHA01}. Thus, from a HFS measurement of a single species
({\it i.e.} H- or Li-like) the QED effects cannot be determined
accurately.

Equation (\ref{eq1}) can also be written as $E^{1s}_{HFS}=C^{1s}+
E^{1s}_{QED}$, where the constant $C^{1s}$ includes everything
except the QED effects. Since the equations for the $(1s)$ and
$(2s)$ states are so similar, it is possible to write the
difference between the two HFS as $\Delta E_{HFS}=E^{2s}_{HFS}-\xi
E^{1s}_{HFS}=E_{non-QED}+E_{QED}$ \cite{SHA01}. The factor $\xi$
only contains non-QED terms and can be calculated to a high
precision \cite{SHA01}. From the difference between the HFS
measurements of H- and Li-like ions of the same isotope, the QED
effects can thus be determined within a few percent. However, this
requires measurements of transition wavelengths with an
experimental resolution of the order of $10^{-6}$.

The transition lifetime $t$ is defined as $t=A^{-1}$ (see {\it
e.g.} \cite{DEM96}). The transition probability $A$, for an $M1$
transition from the excited to the lower hyperfine state, is given
by \cite{BEI00}
\begin{equation}
\label{eq2} A=\frac{4 \alpha (2 \pi \nu)^3 \hbar^2 I \left( 2
\kappa +1 \right)^2}{27m_e^2 c^4 \left( 2I+1 \right)}
\end{equation}
where $\hbar$ is Planck's constant divided by $2 \pi$ and $\kappa$
is related to the electron's angular momentum \cite{BEI00}. From
eq.(\ref{eq2}) it can be seen that $A$ scales with the transition
frequency as $\nu^3$, whereas $\nu$ is proportional to $Z^3$, see
eq.(\ref{eq1}). Therefore, the transition lifetime scales with $Z$
as $t \propto Z^{-9}$ and is roughly of the order of milliseconds
for $Z>70$.

\begin{table}[tb]
\caption{Calculated HFS transition wavelengths ($\lambda$) and
lifetimes ($t$) of the most interesting ion species for systematic
studies. Also shown are the nuclear spin ($I$) and magnetic moment
($\mu$), taken from \cite{FIR98}. The half-lives of these species
are longer than 10 minutes. (The values listed are truncated and
the QED and BW effects are not included.)}
\begin{tabular}{llccccc}
\hline
element & ion & type & $\lambda$ (nm) & $t$ (ms) & $I$ & $\mu$ ($\mu_N$) \\
\hline
& & & & & \\
lead & $^{207}$Pb$^{81+}$ & H-like & 973 & 45 & 1/2 & 0.59 \\
& & & & & \\
bismuth & $^{209}$Bi$^{82+}$ & H-like & 239 & 0.38 & 9/2 & 4.11 \\
 & $^{209}$Bi$^{80+}$ & Li-like & 1469 & 87 &  & \\
& & & & & \\
protactinium & $^{231}$Pa$^{90+}$ & H-like & 262 & 0.64 & 3/2 & 2.01 \\
 & $^{231}$Pa$^{88+}$ & Li-like & 1511 & 123 &  & \\
 & & & & & \\
\hline
& & & & & \\
lead \cite{ROT89} & $^{207}$Pb$^+$ & P$_{3/2}$ - P$_{1/2}$ & 710 & 41 & 1/2 & 0.59 \\
chlorine \cite{BOW60} & $^{35}$Cl$^+$ & $^3$P$_2$ - $^1$D$_2$ & 858 & - & 3/2 & 0.82 \\
 &  & $^3$P$_1$ - $^1$D$_2$ & 913 & - &  &  \\
argon \cite{PRI84} & $^{37}$Ar$^{2+}$ & $^3$P$_2$ - $^1$D$_2$ & 714 & - & 7/2 & 1.3 \\
 &  & $^3$P$_1$ - $^1$D$_2$ & 775 & - &  &  \\
 & & & & & \\
\hline
\end{tabular}
\label{tab1}
\end{table}

In table \ref{tab1} the calculated transition wavelengths
($\lambda$) and lifetimes ($t$), together with their corresponding
$I$ and $\mu$ values, of the most interesting species for our
laser spectroscopy studies are listed. (The QED and BW effects are
not taken into account.) The half-lives of these species exceed 10
minutes, which corresponds to the minimum time required for a
measurement. Although the wavelengths span a broad range, roughly
from 200 to 1600 nm, these transitions are still accessible with
standard laser systems. The three species (Pb \cite{ROT89}, Cl
\cite{BOW60} and Ar \cite{PRI84}) at the bottom of the table are
considered as candidates for pilot experiments. They are singly
charged ions, which are easily produced, have $M1$ transitions at
convenient wavelengths, and can be used to test the laser
spectroscopy part of the experiment. A measurement of the HFS in
$^{207}$Pb$^+$ is of special interest, because it will be possible
to extract the value of $\mu$. Currently two different values
exist, which unfortunately leads to a 2\% difference in the HFS
calculations \cite{SEE98}.

In principle, similar experiments could be carried out with
metastable hafnium ($^{180}$Hf, level energy 1141 keV, half-life
5.5 h \cite{FIR98}). For H-like hafnium, the transition values are
$\lambda=217$ nm and $t=0.25$ ms. For the Li-like ion,
$\lambda=1434$ nm and $t=72$ ms are obtained. The difficulty with
this isotope is that its nucleus is in an excited state, which is
difficult to produce.

\begin{figure}[tb]
\centerline{\includegraphics[width=7.5cm]{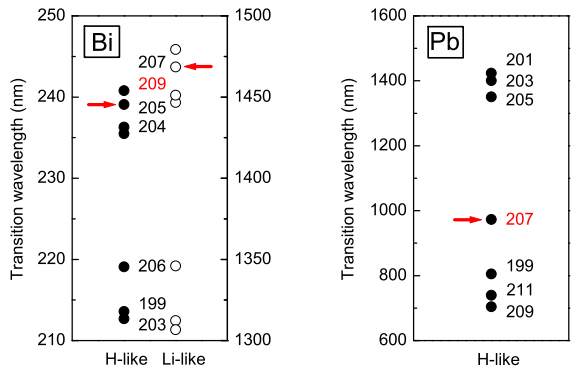}}
\caption{Calculated transition wavelengths for H-like Pb and Bi
isotopes (full circles), and Li-like Bi isotopes (open circles).
Only isotopes with half-lives exceeding 10 minutes are shown. The
small arrows indicate the stable isotopes, the numbers are the
masses in $u$. (The QED and BW effects are not included.)}
\label{fig1}
\end{figure}

Figure \ref{fig1} shows the calculated transition wavelengths of
all H-like lead, and all H- and Li-like bismuth isotopes with
half-lives exceeding 10 minutes. (The QED and BW effects are not
included.) The isotopes are labelled by their corresponding atomic
mass units (in $u$) and the stable isotopes ($^{207}$Pb and
$^{209}$Bi) are indicated by the small arrows. For Pb, only the
H-like isotopes are accessible with standard lasers, because the
transition wavelengths of Li-like isotopes are much longer than
1600 nm. For Bi, many isotopes of both ion species are accessible,
although their transition wavelengths differ considerably.

From Fig.\ref{fig1} it is clear that both elements offer many
candidates for laser spectroscopy measurements of ground state HFS
and that bismuth, in particular, allows for a systematic study of
the (higher order) QED effects at high fields. Furthermore, a
systematic study of different isotopes of the same species, for
example a study of the H-like Pb isotopes, will make it possible
to study trends in nuclear properties across a range of isotopes.

There already exist two previous measurements of the $2s$ ground
state HFS in $^{209}$Bi. A direct measurement \cite{BOR00} was
carried out using the ESR at GSI (Darmstadt), but unfortunately no
resonance was found at the predicted value of $\approx 1554$ nm
\cite{SHA01}. An indirect measurement \cite{BEI98} was performed
in an electron beam ion trap (EBIT) and yielded a value of
$\approx 1512$ nm, but the error in the measurement was rather
large ($\approx 50$ nm). In the ESR the ions have relativistic
velocities ($\approx 200$ MeV/u), which are used to shift the
transition wavelength to a lower value ($\approx 532$ nm), and the
transitions are Doppler-broadened ($\approx 40$ GHz). In the EBIT
the ions have temperatures of several hundreds of eV ($\sim 10^6$
K), which lead to considerable Doppler broadening ($\approx 10$
GHz). The resolution obtained in previous measurements at the ESR
is of the order of $10^{-4}$, whereas that of the EBIT measurement
is of the order of $10^{-2}$.

\section{Experiment overview}
A detailed description of the proposed experiments, as well as a
treatment of the techniques used, can be found elsewhere
\cite{WIN05,VOG05}. Briefly, an externally produced bunch of
roughly $10^5$ HCI at an energy of a few keV is loaded into a
cylindrical open-endcap Penning trap \cite{GAB89} on axis, {\it
i.e.} along the magnetic field lines. Electron capture
(neutralisation) by collisions is strongly reduced by operating
the trap at cryogenic temperatures under UHV conditions. The HCI
are captured in flight, confined, cooled by `resistive cooling'
\cite{WIN75} and radially compressed by a `rotating wall'
\cite{ITA98} technique. After these steps a cold and dense ion
cloud is obtained. The spectroscopy laser enters the trap axially
through an open-endcap and will fully irradiate the ion cloud. The
fluorescence from the excited HCI is detected perpendicular to the
cooled axial motion (trap axis) through segmented ring electrodes,
which are covered by a highly-transparent copper mesh. (The ring
is segmented for the rotating wall technique.)

The above mentioned transition lifetimes imply that, for a
detection efficiency of $\sim 10^{-3}$, acceptable fluorescence
rates, up to a few thousand counts per second, from $M1$
transitions can be expected from a ($\sim3$ mm diameter) cloud of
$10^5$ ions \cite{WIN05,VOG05}. Confining the HCI in a trap, and
cooling and compressing the cloud, will thus enable fluorescence
detection and ensure long interrogation times by the laser.

However, due to the high density of HCI in the cloud, space charge
effects will play a role and will lead to shifts of the motional
frequencies of the trapped ions. We have studied this effect in
detail and understand the corresponding frequency shifts well
\cite{WIN06}. Since these shifts are fairly small, the (frequency
dependent) cooling and compression techniques can still be
applied.

The HCI also need to be strongly cooled to reduce Doppler
broadening of the transitions. This will be achieved by resistive
cooling of the (axial) ion motion in the trap. For example, for
the $F=1 \rightarrow F=0$ transition in $^{207}$Pb$^{81+}$ at $\nu
\approx 3 \times 10^{14}$ Hz, the Doppler broadened linewidth at a
temperature of 4 K is $\Delta \nu_D \approx 3 \times 10^7$ Hz. The
anticipated resolution is therefore of the order of
$10^{7}/10^{14}=10^{-7}$. This is three orders of magnitude better
than any previous measurement, see {\it e.g.} \cite{SEE98,KLA94},
and good enough to measure the QED effects within a few percent.

\section{Acknowledgments}
This work is supported by the European Commission within the
framework of the HITRAP project (HPRI-CT-2001-50036). W.N.
acknowledges funding by the Helmholtz Association (VH-NG-148).

\end{document}